\documentclass[aps,showpacs,preprintnumbers,amsmath, amssymb]{revtex4}

\usepackage{float}
\usepackage{graphics,epsfig}
\usepackage{graphicx}
\usepackage{dcolumn}
\usepackage{bm}

\begin{document}
\baselineskip=0.8 cm

\title{{\bf No scalar hair theorem for neutral Neumann stars: static massive scalar fields nonminimally coupled to gravity}}
\author{Yan Peng$^{1}$\footnote{yanpengphy@163.com}}
\affiliation{\\$^{1}$ School of Mathematical Sciences, Qufu Normal University, Qufu, Shandong 273165, China}

\vspace*{0.2cm}
\begin{abstract}
\baselineskip=0.6 cm
\begin{center}
{\bf Abstract}
\end{center}

In a recent paper, Hod proved that spherically symmetric
Dirichlet reflecting compact stars cannot support static
nonminimally coupled scalar fields.
In the present paper, we study the validity of no hair theorems
for compact stars with Neumann surface boundary conditions.
We find that Neumann compact stars cannot support static massive
scalar field hairs with a generic dimensionless nonminimal coupling parameter.

\end{abstract}

\maketitle
\newpage
\vspace*{0.2cm}

\section{Introduction}

The classical no hair theorem \cite{Bekenstein}-\cite{JBN}
plays an important role in the development of black hole theories.
It states an intriguing property that asymptotically flat black holes cannot support
static scalar fields, for recent progress
see references \cite{mr1}-\cite{CAHE} and reviews \cite{Bekenstein-1,CAR}.
It was usually believed that this no hair property is due to the existence of black hole
absorbing horizons.

However, no scalar hair
behavior also appears in the horizonless spacetime.
It was firstly proved that asymptotically flat
neutral Dirichlet reflecting horizonless compact stars cannot
support massive scalar field hairs \cite{Hod-6}.
In the asymptotically dS gravity, massive scalar,
vector and tensor fields also cannot exist outside
neutral horizonless Dirichlet
reflecting compact stars \cite{Bhattacharjee}.
Then whether no hair theorem exists in the charged horizonless gravity
is still an question to be answered.
In fact, it was shown that static scalar fields cannot condense
outside charged reflecting shells of large radii \cite{Hod-8,Hod-9,Yan Peng-1}.
Large charged reflecting stars also cannot support
static scalar field hairs
\cite{Hod-10,Yan Peng-2,Yan Peng-3,Yan Peng-4,Yan Peng-5,YP1,YP2,YP3}.
Moreover, it was found that scalar fields cannot exist outside compact stars with
Neumann surface boundary conditions \cite{Yan Peng-6,Yan Peng-7}.

All front no scalar hair theorems only consider minimally
coupled scalar field hairs. Interestingly, no
scalar hair theorem also holds with nonminimal field-curvature
couplings \cite{ADB,AS,Hod-11,Hod-12}. Considering a nonminimal coupling parameter
$\xi$, compact stars with generic boundary conditions
can rule out the existence of scalar hairs in ranges
$\xi<0$ and $\xi>\frac{1}{4}$ and
compact stars with Dirichlet reflecting
boundary conditions can also rule out exterior scalar hair
for $0\leqslant\xi\leqslant\frac{1}{4}$ \cite{Hod-13,Hod-14,Hod-15}.
So it is interesting to study no scalar hair properties
with other boundary conditions in the range $0\leqslant \xi\leqslant\frac{1}{4}$.
In this work, we plan to investigate the
no nonminimal scalar hair behavior in the
background of horizonless compact stars with Neumann boundary conditions.

The rest of this work is as follows.
We construct the model of static
scalar fields nonminimally coupled to gravity in the
background of Neumann compact stars.
We find that no scalar hair theorem
holds for generic coupling parameters.
At last, we summarize the main results.

\section{No hair theorem for scalar fields outside Neumann stars}

We study the model of massive scalar fields
nonminimally coupled to the compact star gravity.
And the asymptotically flat spherically symmetric spacetime reads \cite{ADB,AS,Hod-11,Hod-12,Hod-13}
\begin{eqnarray}\label{AdSBH}
ds^{2}&=&-e^{\nu}dt^{2}+e^{\lambda}dr^{2}+r^{2}(d\theta^2+sin^{2}\theta d\phi^{2}).
\end{eqnarray}
The functions $\nu$ and $\lambda$
only depend on the radial coordinate r.
We define the radial coordinate $r=r_{s}$ as the star radius.
Asymptotic flatness of the spacetime requires
the behaviors \cite{ADB}
\begin{eqnarray}\label{AdSBH}
\nu(r\rightarrow \infty)\backsim O(\frac{1}{r}),~~~~~ \lambda(r\rightarrow \infty)\backsim O(\frac{1}{r}).
\end{eqnarray}

The Lagrange density describing scalar fields
nonminimally coupled to gravity is \cite{Hod-12,Hod-13}
\begin{eqnarray}\label{lagrange-1}
\mathcal{L}=R-\xi R\psi^2-|\nabla_{\alpha} \psi|^{2}-\mu^{2}\psi^{2},
\end{eqnarray}
where $\psi(r)$ is the scalar field with mass $\mu$.
We label R as scalar Ricci curvature of the spacetime.
Asymptotic flatness of the spacetime yields
\begin{eqnarray}\label{AdSBH}
R(r\rightarrow \infty)\rightarrow 0.
\end{eqnarray}
The dimensionless parameter $\xi$ describes
the nonminimal coupling strength between scalar fields and curvature.
Hod proved that compact stars with generic boundary conditions
can rule out the existence of scalar hairs in ranges
$\xi<0$ and $\xi>\frac{1}{4}$ \cite{Hod-13}.
So it remains to study no hair theorem for parameters satisfying
\begin{eqnarray}\label{AdSBH}
0~\leqslant ~\xi~\leqslant~\frac{1}{4}.
\end{eqnarray}

The scalar field equation is
\begin{eqnarray}\label{BHg}
\psi''+(\frac{2}{r}+\frac{\nu'}{2}-\frac{\lambda'}{2})\psi'-(\mu^2+\xi R)e^{\lambda}\psi=0.
\end{eqnarray}

Around the infinity, the scalar field asymptotically behaves
in the form
\begin{eqnarray}\label{AdSBH}
\psi\sim A\cdot\frac{1}{r}e^{-\mu r}+B\cdot\frac{1}{r}e^{\mu r},
\end{eqnarray}
where A and B are integral constants.
The physical solution requires $B=0$ \cite{Hod-12}.
At the star surface, we take the Neumann boundary condition.
It yields boundary conditions
\begin{eqnarray}\label{InfBH}
&&\psi'(r_{s})=0,~~~~~~~~~\psi(\infty)=0.
\end{eqnarray}

The Ricci scalar curvature is
\begin{eqnarray}\label{BHg}
R=-\frac{8\pi}{1-8\pi\xi^2\psi^2}\{e^{-\lambda}[\xi(\frac{12}{r}+3\nu'-3\lambda')\psi\psi'+6\xi\psi\psi''+(6\xi-1)(\psi')^2]-2\mu^2\psi^2\}.
\end{eqnarray}

Substituting (9) into the scalar field equation (6), we arrive at
\begin{eqnarray}\label{BHg}
F\cdot\psi''+[F\cdot(\frac{2}{r}+\frac{\nu'}{2}-\frac{\lambda'}{2})
+8\pi\xi(6\xi-1)\psi\psi']\psi'-\mu^2e^{\lambda}(1+8\pi\xi\psi^2)\psi=0,
\end{eqnarray}
where $F(r,\xi)=1+8\pi\xi(6\xi-1)\psi^2$.

We divide the proof of no hair theorem into three cases
\begin{eqnarray}\label{BHg}
\psi(r_{s})=0,~~ \psi(r_{s})>0 ~~and~~ \psi(r_{s})<0.
\end{eqnarray}
It is known that the nontrivial scalar field cannot exist for $\psi(r_{s})=0$ \cite{Hod-13}.
Considering the symmetry $\psi\rightarrow -\psi$ of equation (10),
it remains to prove no hair theorem for
\begin{eqnarray}\label{BHg}
\psi(r_{s})>0.
\end{eqnarray}

In the range $0\leqslant\xi\leqslant \frac{1}{4}$, there is $F(r,\xi)\geqslant0$,
which is important in the following analysis \cite{Hod-13}.
The proof of $F(r,\xi)\geqslant0$ is as follows.
In the range $\frac{1}{6}\leqslant\xi\leqslant \frac{1}{4}$, there is $F(r,\xi)\geqslant0$.
For $0\leqslant\xi<\frac{1}{6}$, $F(r,\xi)$ cannot switch signs. Otherwise, $F(r,\xi)$ vanishes at some point $r_{0}$.
And at this point $r=r_{0}$, there is the relation
\begin{eqnarray}\label{BHg}
8\pi\xi(6\xi-1)\psi'^2=\mu^2e^{\lambda}(1+8\pi\xi\psi^2).
\end{eqnarray}
In the regime $0\leqslant\xi<\frac{1}{6}$, the functional expression
on the left side of (13) is non-positive whereas the functional
expression on the right side of (13) is positive. One therefore
deduces that the radial function $F(r,\xi)$ cannot switch signs.
At the infinity, $F(r,\xi)$ behaves as
\begin{eqnarray}\label{BHg}
F(r\rightarrow \infty,\xi)\rightarrow 1+8\pi\xi(6\xi-1)\psi(\infty)^2=1>0.
\end{eqnarray}
So we find $F(r,\xi)>0$ in the case of $0\leqslant\xi<\frac{1}{6}$.
As a summary, for $0\leqslant\xi\leqslant \frac{1}{4}$, there is the relation
\begin{eqnarray}\label{BHg}
F(r,\xi)\geqslant0.
\end{eqnarray}

We divide the analysis into two cases
\begin{eqnarray}\label{BHg}
\psi''(r_{s})\leqslant 0 ~~~~and~~~~ \psi''(r_{s})>0.
\end{eqnarray}

In the case of $\psi''(r_{s})\leqslant 0$, we obtain following relations at the star surface $r=r_{s}$ as
\begin{eqnarray}\label{InfBH}
\{ \psi^2>0,~~~~\psi'=0~~~~and~~~~\psi \psi''\leqslant0\}~~~~for~~~~r=r_{s}.
\end{eqnarray}

At the star radius $r_{s}$, relations (15), (17) and
$0\leqslant\xi\leqslant \frac{1}{4}$ give the characteristic inequality
\begin{eqnarray}\label{BHg}
F\cdot\psi\psi''+[F\cdot(\frac{2}{r}+\frac{\nu'}{2}-\frac{\lambda'}{2})
+8\pi\xi(6\xi-1)\psi\psi']\psi\psi'-\mu^2e^{\lambda}(1+8\pi\xi\psi^2)\psi^2<0,
\end{eqnarray}
which is in contradiction with equation (10).

In another case of $\psi''(r_{s})>0$, with the condition $\psi'(r_{s})=0$,
there is $\psi'(r)>0$ around $r_{s}$.
With increase of the radial coordinate, the scalar field firstly becomes
more positive and finally approaches zero at the infinity.
In this case, there is at least one positive maximum extremum point $r=r_{peak}$
between the star surface $r_{s}$ and the infinity boundary.
At this extremum point, the scalar field is characterized by following relations
\begin{eqnarray}\label{InfBH}
\{ \psi^2>0,~~~~\psi'=0~~~~and~~~~\psi \psi''\leqslant0\}~~~~for~~~~r=r_{peak}.
\end{eqnarray}

At this extremum point $r=r_{peak}$, relations (15), (19) and
$0\leqslant\xi\leqslant \frac{1}{4}$ lead to the inequality
\begin{eqnarray}\label{BHg}
F\cdot\psi\psi''+[F\cdot(\frac{2}{r}+\frac{\nu'}{2}-\frac{\lambda'}{2})
+8\pi\xi(6\xi-1)\psi\psi']\psi\psi'-\mu^2e^{\lambda}(1+8\pi\xi\psi^2)\psi^2<0.
\end{eqnarray}
It can be easily seen that relations (18) and (20) are in contradiction with equation (10).
So nontrivial scalar field solution of equation (10) cannot exist.
Here we prove no nonminimally coupled scalar hair theorem for
$0\leqslant\xi\leqslant \frac{1}{4}$.
Also considering known results that compact stars with generic boundary conditions
cannot support scalar hairs in ranges
$\xi<0$ and $\xi>\frac{1}{4}$ \cite{Hod-13}, we conclude that
scalar hair cannot form outside regular neutral Neumann stars
for any coupling parameter $\xi$.

\section{Conclusions}

In the background of Neumann stars, we studied no hair theorem for static massive scalar fields
nonminimally coupled to the spherically symmetric asymptotically flat horizonless gravity.
We considered the field-curvature coupling and
scalar fields' backreaction on the background.
We obtained the characteristic inequalities (18) at the star surface
and (20) at extremum points,
which are in contradiction with the scalar field equation (10).
It means that there is no nontrivial scalar field solution.
So we concluded that asymptotically flat spherically symmetric regular Neumann
stars cannot support the existence of exterior massive scalar field hairs
for generic nonminimal coupling parameters.

\begin{acknowledgments}

We would like to thank the anonymous referee for the constructive suggestions to improve the manuscript.
This work was supported by the Shandong Provincial Natural Science Foundation of China under Grant
No. ZR2018QA008. This work was also supported by a grant from Qufu Normal University of China under Grant
No. xkjjc201906.

\end{acknowledgments}


\begin{thebibliography}{99}




\bibitem{Bekenstein}
J. D. Bekenstein, Transcendence of the law of baryon-number conservation in black hole physics, Phys. Rev. Lett. 28, 452 (1972).




\bibitem{Chase}
J. E. Chase, Event horizons in Static Scalar-Vacuum Space-Times, Commun. Math. Phys. 19, 276 (1970).



\bibitem{C. Teitelboim}
C. Teitelboim, Nonmeasurability of the baryon number of a black-hole, Lett. Nuovo Cimento 3, 326 (1972).





\bibitem{Ruffini-1}
R. Ruffini and J. A. Wheeler, Introducing the black hole, Phys. Today 24, 30 (1971).




\bibitem{W.K.H}
W.K.H. Panofsky, Needs Versus Means In High-energy Physics, Phys. Today 33(1980)24-33.




\bibitem{MH1}
M. Heusler, A No hair theorem for selfgravitating nonlinear sigma models, J. Math. Phys. 33(1992)3497-3502.



\bibitem{MH2}
Markus Heusler, A Mass bound for spherically symmetric black hole space-times, Class. Quant. Grav. 12(1995)779-790.


\bibitem{JBN}
J.D. Bekenstein, Novel ``no-scalar-hair'' theorem for black holes, Phys. Rev. D 51(1995)no.12,R6608.







\bibitem{mr1}
D. N$\acute{u}\tilde{n}$ez, H. Quevedo, and D. Sudarsky,Black Holes Have No Short Hair,Phys. Rev. Lett.
76, 571 (1996).



\bibitem{mr2}
S. Hod,Hairy Black Holes and Null Circular Geodesics,Phys. Rev. D 84,124030(2011).






\bibitem{e5}
Pallab Basu, Chethan Krishnan, P.N.Bala Subramanian, Hairy black holes in a box, JHEP 11(2016)041.




\bibitem{Hod-1}
S. Hod, Stationary Scalar Clouds Around Rotating Black Holes, Phys. Rev. D 86, 104026 (2012).






\bibitem{Herdeiro-1}
C. A. R. Herdeiro and E. Radu, Kerr black holes with scalar hair, Phys. Rev. Lett. 112, 221101(2014).


\bibitem{Benone}
C. L. Benone, L. C. B. Crispino, C. Herdeiro, and E. Radu, Kerr-Newman scalar clouds, Phys.
Rev. D 90, 104024 (2014).






\bibitem{NSHB}
Yan Peng, Hair mass bound in the black hole with non-zero cosmological constants,
Physical Review D 98(2018)104041.



\bibitem{HEB}
Yan Peng, Hair distributions in noncommutative Einstein-Born-Infeld black holes,
Nucl. Phys. B 941(2019)1-10.



\bibitem{NGT}
Yan Peng,No-Go Theorem for Asymptotically AdS Static Regular Boson Stars,arXiv:1902.06508[gr-qc].





\bibitem{Degollado}
J. C. Degollado and C. A. R. Herdeiro, Stationary scalar configurations around extremal charged black holes,
Gen. Rel. Grav. 45, 2483 (2013).


\bibitem{Cunha}
P. V. P. Cunha, C. A. R. Herdeiro, E. Radu, and H. F. R¡äunarsson, Shadows of Kerr black holes with scalar hair, Phys.
Rev. Lett. 115, 211102(2015).






\bibitem{Brihaye}
Y. Brihaye, C. Herdeiro, and E. Radu, Inside black holes with synchronized hair, Phys. Lett. B 760, 279 (2016).



\bibitem{e1}
S. A. Hartnoll, C. P. Herzog and G. T. Horowitz, Holographic Superconductors, JHEP 0812(2008)015.






\bibitem{bl1}
Nicolas Sanchis-Gual,Juan Carlos Degollado,Pedro J.Montero,Jos A. Font,Carlos Herdeiro, Explosion and final
state of an unstable Reissner-Nordstr$\ddot{o}$m black hole, Phys. Rev. Lett. 116(2016)141101.


\bibitem{bl2}
Sam R Dolan,Supakchai Ponglertsakul,Elizabeth Winstanley, Stability of black holes in Einstein-charged scalar
field theory in a cavity, Phys. Rev. D 92(2015)124047.





\bibitem{CW}
Chen Wu, Renli Xu, Decay of massive scalar field in a black hole background immersed in magnetic field,
Eur. Phys. J. C 75(2015) no.8,391.


\bibitem{Marek Rogatko1}
Marek Rogatko, Uniqueness of higher-dimensional phantom field wormholes, Phys. Rev. D 97(2018)no.2,024001.


\bibitem{Marek Rogatko2}
Marek Rogatko, Uniqueness of higher-dimensional Einstein-Maxwell-phantom dilaton field wormholes, Phys. Rev. D 97(2018)no.6,064023.



\bibitem{PWHH}
Peng Wang, Houwen Wu, Haitang Yang,Thermodynamics and Phase Transition of a Nonlinear Electrodynamics Black Hole in a Cavity, arXiv:1901.06216[gr-qc].

\bibitem{WBL}
Ming Zhang, Shan-Zhong Han, Jie Jiang, Wen-Biao Liu,Circular orbit of a test particle and phase transition of a black hole,Phys. Rev. D 99(2019) no.6,065016.


\bibitem{CAHE}
Carlos A.R. Herdeiro, Eugen Radu, Black hole scalarization from the breakdown of scale invariance, Phys. Rev. D 99(2019)no.8,084039.












\bibitem{Bekenstein-1}
J. D. Bekenstein, Black hole hair: 25-years after, arXiv:gr-qc/9605059.






\bibitem{CAR}
Carlos A. R. Herdeiro, Eugen Radu,Asymptotically flat black holes with scalar hair: a review, Int.J.Mod.Phys.D
24(2015)09,1542014.








\bibitem{Hod-6}
S.Hod, No-scalar-hair theorem for spherically symmetric reflecting stars, Physical Review D 94, 104073 (2016).





\bibitem{Bhattacharjee}
Srijit Bhattacharjee, Sudipta Sarkar, No-hair theorems for a static and stationary reflecting star, Physical Review D 95, 084027 (2017).









\bibitem{Hod-8}
S.Hod, Charged massive scalar field configurations supported by a spherically symmetric
charged reflecting shell,Physics Letters B 763, 275 (2016).



\bibitem{Hod-9}
S.Hod, Marginally bound resonances of charged massive scalar fields in the background of
a charged reflecting shell, Physics Letters B 768(2017)97-102.


\bibitem{Yan Peng-1}
Yan Peng, Bin Wang, Yunqi Liu, Scalar field condensation behaviors around reflecting shells in
Anti-de Sitter spacetimes, Eur.Phys.J. C 78 (2018) no.8, 680.


\bibitem{Yan Peng-2}
Yan Peng, Scalar field configurations supported by charged compact reflecting stars in a curved spacetime,
Physics Letters B 780(2018)144-148.



\bibitem{Hod-10}
Shahar Hod, Charged reflecting stars supporting charged massive scalar field configurations, European Physical
Journal C 78, 173 (2017).











\bibitem{Yan Peng-3}
Yan Peng, Static scalar field condensation in regular asymptotically AdS reflecting star backgrounds,
Phys. Lett. B 782(2018)717-722.



	
\bibitem{Yan Peng-4}
Yan Peng, On instabilities of scalar hairy regular compact reflecting stars, JHEP 10(2018)185.





\bibitem{Yan Peng-5}
Yan Peng, Hair formation in the background of noncommutative reflecting stars, Nucl.Phys. B 938(2019)143-153.



\bibitem{YP1}
Yan Peng, No hair theorem for spherically symmetric regular compact stars with Dirichlet boundary conditions, Phys. Lett. B 792(2019)156-159.



\bibitem{YP2}
Yan Peng, Large regular reflecting stars have no scalar field hair, Eur. Phys. J. C 79(2019)no.4,309.




\bibitem{YP3}
Yan Peng, No hair theorem for massless scalar fields outside asymptotically flat horizonless reflecting compact stars, arXiv:1904.00911[gr-qc].










\bibitem{Yan Peng-6}
Yan Peng, Scalar condensation behaviors around regular Neumann reflecting stars, Nucl.Phys. B 934 (2018) 459-465.




\bibitem{Yan Peng-7}
Yan Peng, No hair theorem for bound-state massless static scalar fields outside horizonless Neumann compact stars,
arXiv:1905.04541[gr-qc].



\bibitem{ADB}
Avraham E. Mayo, Jacob D. Bekenstein, No hair for spherical black holes: Charged and
nonminimally coupled scalar field with selfinteraction, Phys. Rev. D 54(1996)5059-5069.



\bibitem{AS}
Alberto Saa, Searching for nonminimally coupled scalar hairs, Phys. Rev. D 53(1996)7377-7380.











\bibitem{Hod-11}
S.Hod, No nonminimally coupled massless scalar hair for spherically
symmetric neutral black holes, Physics Letters B 771(2017)521-523.




\bibitem{Hod-12}
S.Hod, No hair for spherically symmetric neutral black holes: Nonminimally
coupled massive scalar fields, Physical Review D 96,124037(2017).












\bibitem{Hod-13}
S.Hod, No hair for spherically symmetric neutral reflecting stars:
Nonminimally coupled massive scalar fields, Physics Letters B 773(2017)208-212.







\bibitem{Hod-14}
S.Hod, No nonminimally coupled massless scalar hair for spherically symmetric neutral reflecting stars,Physical Review D 96, 024019 (2017).





\bibitem{Hod-15}
S.Hod,No-go theorem for static boson stars,Physics Letters B 778(2018)239-241.













\end{thebibliography}
\end{document}